# Magic-free coexisting photonic and phononic moiré flat bands

**Authors**: Ziming Chen[1], Kaiyu Cui[1,2,*], Ning Wu[1], Shuyuan Li[1], Chenxuan Wang[1], Xue Feng[1], Fang Liu[1], Wei Zhang[1], Hao Sun[1], Yongzhuo Li[1], and Yidong Huang[1]

**Affiliations:**

[1] Department of Electronic Engineering, Tsinghua University, Beijing 100084, P. R. China

[2] Beijing National Research Center for Information Science and Technology (BNRist), Beijing 100084, P. R. China

[*] Corresponding author: kaiyucui@tsinghua.edu.cn

**Abstract:** Moiré flat bands enhance localization and interactions through suppressed group velocity, but existing approaches largely target a single physical field because distinct excitations generally require different, finely tuned magic configurations. Here we introduce a flat-band mechanism based on strong diffractive hybridization among moiré-folded bands. Period-mismatched modulations open distinct coupling channels whose hybridization renormalizes the band dispersion. An effective Hamiltonian shows that increasing the diffractive coupling progressively suppresses the group velocity, driving the system toward a flat-band regime without field-specific magic configurations. This coupling-induced mechanism enables band flattening across distinct physical excitations. We demonstrate this mechanism in a single-layer moiré optomechanical crystal, where photonic and phononic flat bands are simultaneously realized, and their localized modes and optomechanical interaction are experimentally observed. Beyond photonic and phononic systems, this mechanism may extend to other wave and quasiparticle platforms, providing a general route to co-localizing and coupling distinct physical fields in moiré systems.

## 1. Introduction

Moiré materials[1, 2, 3, 4, 5, 6] have emerged as a versatile platform for exploring exotic phenomena. The band structure of moiré materials can be tuned by adjusting interlayer coupling through changing twist angle or lattice mismatch, which dramatically modifies the properties of the systems. At certain geometric conditions such as the so-called magic angle[1, 7, 8, 9], flat bands appear, characterized by near-zero group velocity and high density of states (DoS). In electronic systems, such flat bands can give rise to enhanced interactions[10, 11], correlated insulating phases[7, 12], superconductivity[1, 13], and strongly correlated topological states[14, 15], where electron-electron interaction energy becomes comparable to or larger than the electronic kinetic-energy scale near the Fermi level. Beyond electronic systems, the underlying principles can be extended [3] to photonic[2, 16, 17, 18, 19, 20, 21, 22, 23, 24], acoustic[6, 25, 26, 27, 28], thermal diffusion[29], and hydrodynamic systems[5], where flat bands have been exploited to confine optical modes for high-quality lasers[2, 18] or to achieve unidirectional acoustic canalization[27].

Despite growing interest in moiré lattices across different physical systems, most studies have focused on a single type of physical field, leaving the simultaneous manipulation of multiple types of physical fields by flat bands largely unexplored. Flat bands can confine multiple types of physical fields within the same structure simultaneously, and this co-localization can enhance their interactions through their high density of states (DoS). The central challenge here is not merely to create flat bands, but to do so simultaneously for fields with very different dispersion relations. In multilayer moiré systems[1, 8, 9, 21, 30], flat bands typically rely on finely tuned magic configurations (a set of discrete geometric conditions for flat bands), making it difficult to simultaneously achieve flat bands for fields with distinct dispersions, which hinders the co-localization and control of multiple types of fields in a shared moiré lattice.

Cavity optomechanics[31] studies the interaction between optical and mechanical fields, leading to phenomena such as phonon lasing[32], macroscopic mechanical ground states[33],etc. Optomechanical devices coherently couple optical and mechanical fields, with broad applications such as wavelength conversion[34], quantum memory[35], and high-precision sensing[36]. While photonic and phononic flat bands can localize optical and mechanical modes and possibly enhance optomechanical coupling with increased spatial overlap, their simultaneous realization within a single platform has remained largely unexplored.

Motivated by this challenge, we theoretically and experimentally demonstrate a one-dimensional single-layer moiré optomechanical crystal formed by superimposing two periodic modulations with slightly mismatched periods. The resulting structure supports multiple optical and mechanical flat bands. We show that, unlike multilayer moiré systems that rely on finely tuned magic configurations, band flattening in this single-layer moiré lattice is governed primarily by strong inter-band coupling among multiple bands of the given physical fields. As the coupling strength increases, the group velocity is progressively suppressed without repeated zero crossings, which is a monotonic response that removes the need to fine-tune a magic configuration for each field. This field-independent mechanism is confirmed by finite element method (FEM) simulations of both optical and mechanical modes, providing a generic route to their simultaneous localization within the same structure.

Unlike defect-based optomechanical crystal cavities[31, 37, 38, 39, 40, 41], our design provides an alternative route to optomechanical confinement that does not rely on bandgap or defect engineering for defect cavities. Beyond this, the multiple flat bands in our moiré optomechanical crystal enable simultaneous manipulation of different physical fields and multiple modes, providing not only a broad operating frequency range[42] of the devices, but also complex coupling networks between optical and mechanical modes in a compact system for exploring such as multimode optomechanics[43, 44, 45] and topological physics[46, 47]. More broadly, our work provides a compact platform for multi-field moiré engineering, in which distinct physical fields can be co-localized, co-engineered, and mutually coupled. This architecture not only realizes a new type of optomechanical cavity but also offers a versatile framework for investigating interactions between multiple physical fields[24, 26, 48, 49, 50, 51].

## 2. Design of one-dimensional moiré optomechanical crystal

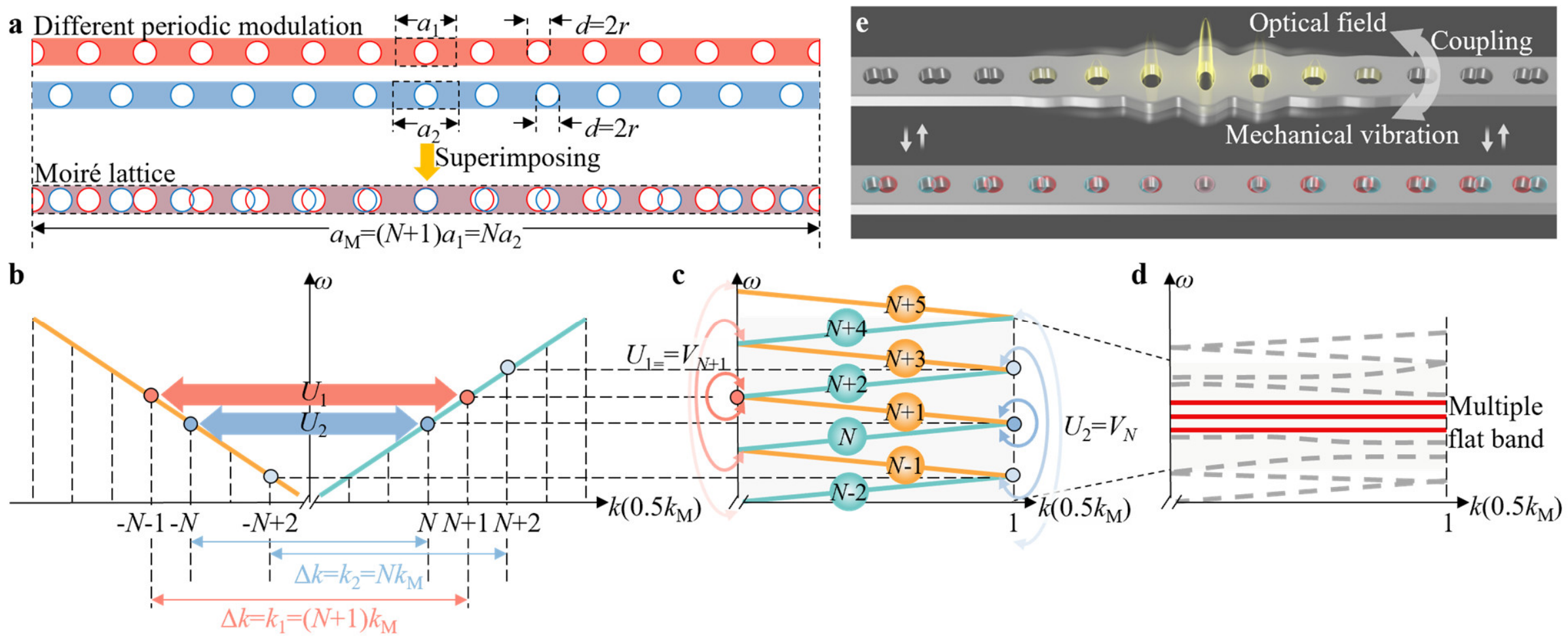

**Fig. 1. Formation of moiré lattice and occurrence of flat band in a moiré optomechanical crystal. a.** Schematic illustration of the formation of the moiré optomechanical crystal. Top and middle panels: Two original nanobeams with mismatched periods but identical air-hole radii. The different periods are shown in the dashed boxes. Bottom panel: Superposition of the two sets of periodic air holes onto the same nanobeam forms the moiré lattice. Red and blue dashed circles mark the air holes from the two original nanobeams, respectively, illustrating the origin of the moiré modulation. **b.** Illustration of forward–backward mode coupling in the two original nanobeams, according to quasi-momentum conservation $k - p = Nk_n$, where $k_1$, $k_2$ and $k_M$ correspond to the reciprocal lattice vector with $k_1 = \frac{2\pi}{a_1} = i_1\frac{2\pi}{a_M} = (N+1)\frac{2\pi}{a_M} = (N+1)k_M$ and $k_2 = \frac{2\pi}{a_2} = i_2\frac{2\pi}{a_M} = N\frac{2\pi}{a_M} = Nk_M$ illustrated with arrows below the $k$ axis. Red and blue arrows represent diffractive couplings associated with the modulation period of each original nanobeam in a, with the same color. **c** and **d.** Forward–backward mode coupling in the moiré lattice illustrated in a. Red and blue arrows have the same meaning as in b, while the degree of transparency indicates weaker splitting where frequency differences are larger. The interplay of the two kinds of diffraction coupling leads to multiple flat bands at some coupling strength illustrated by red lines in d. The shaded region in d corresponds to the bands in the shaded region in c. **e.** Schematic illustration of the optical and mechanical modes in the moiré optomechanical crystal. On the top: the interaction of localized optical mode (yellow wave packet) and the mechanical mode (vibration characterized by deformation); on the bottom: two mismatched periodic holes colored as in a.

In contrast to multilayer moiré lattices in which flat bands occur at certain magic configurations, we begin with a one-dimensional moiré optomechanical crystal to show the formation of multiple flat bands in a single-layer moiré lattice. As illustrated in Fig. 1a, a unit cell of a one-dimensional moiré optomechanical crystal is formed by superimposing the modulation patterns of two periodic silicon nanobeams. These nanobeams are modulated by central air holes with different periods. The ratio between the two periods is chosen to be rational so that the resulting moiré superlattice remains periodic. For simplicity, we select the periods of the two original nanobeams, denoted as $a_1$ and $a_2$, to satisfy the commensurate condition $a_1 : a_2 = N : (N+1)$, where $N$ is a positive integer. Thus, the period of the moiré superlattice is $a_M = (N+1)a_1 = Na_2$.

Compared with twistronics[1, 7, 8, 9] or multilayer photonic crystals[21, 52, 53], the periodic mismatched patterns in our moiré optomechanical crystal are superimposed onto a single layer of nanobeam. This leads to two important differences: first, there is only one structural layer for forward and backward propagation modes, implying only one unperturbed Hamiltonian formed by a pair of forward and backward modes; second, there exist two types of diffractive couplings for forward-backward coupling originating from two sets of periodic patterns. Consequently, the effective Hamiltonian should differ from those describing multilayer moiré systems[8, 21, 30, 52].

Here we establish an effective Hamiltonian model for the moiré lattice with different periodic patterns superimposed onto the same structure. Firstly, for free forward ($c_+(k)$) and backward ($c_-(k)$) propagating modes in the unperturbed nanobeam with dispersion relation as $\omega_\pm = \omega_\pm(k)$, the effective free Hamiltonian $\mathcal{H}_{\text{free}}$ is

$$\mathcal{H}_{\text{free}} = \omega_+(k)c_+^\dagger(k)c_+(k) + \omega_-(p)c_-^\dagger(p)c_-(p). \tag{1}$$

Here the dispersion relation can be linearized within small wave vector regions as $\omega_\pm = \omega_0 \pm v_0 k$ approximately in a moiré lattice with a large period. The diffractive couplings in the moiré lattice can be considered as the collective contribution from all diffractive couplings in the original periodic mismatched lattice with periods $a_n$. For the *l*-th order diffractive coupling $U_{n,l}$ associated with period $a_n$, the coupling term can be described as $U_{n,l}c_+^\dagger(k)c_-(p)\delta\left(k - p = l\dfrac{2\pi}{a_n}\right) + \text{H.C.}$, where the delta function ensures quasi-momentum conservation. Thus, the total diffraction coupling term can be described as

$$\mathcal{H}_{\text{diff}} = c_+^\dagger(k)c_-(p)\sum_{n,l} U_{n,l}\delta(k-p=lk_n) + \text{H.C.}, \qquad k_n = \frac{2\pi}{a_n} = i_n \frac{2\pi}{a_{\text{M}}} = i_n k_{\text{M}}, \tag{2}$$

where $i_n$ is an integer from the above commensurate condition. Thus, the total effective Hamiltonian is $\mathcal{H}_{\text{eff}} = \mathcal{H}_{\text{free}} + \mathcal{H}_{\text{diff}}$. This diffractive coupling process can be illustrated in Fig. 1b, in which diffractive coupling is significant when the frequencies of the two modes are close, inducing band splitting. The diffractive coupling term (2) can be rearranged according to the period of moiré lattice $a_{\text{M}}$:

$$\mathcal{H}_{\text{diff}} = c_+^\dagger(k)c_-(p)\sum_{n} V_n\delta(k-p=nk_{\text{M}}) + \text{H.C.}, \qquad k_{\text{M}} = \frac{2\pi}{a_{\text{M}}}, \tag{3}$$

where we group the diffractive terms according to the quasi-momentum difference $(k-p)$ and use $V_n$ to denote the overall contribution $\sum_{n,l} U_{n,l}$ satisfying the quasi-momentum conservation $k-p=nk_M$. This rearranged coupling term has the same form as the forward-backward coupling model for a crystal with period $a_M$. However, as in the moiré optomechanical model, the strength distribution for different diffractive orders is rearranged according to the original periods of the two lattices. For example, coupling strengths are relatively higher at diffractive orders corresponding to $N,(N+1),2N,2(N+1),\ldots$ in our moiré optomechanical crystals. By folding the original dispersion relation according to the moiré period $a_{\text{M}}$, the above Hamiltonian can be rewritten as

$$\mathcal{H}_{\text{eff}} = \sum_{n=1} \frac{1}{2}\omega_n(k)c_n^\dagger(k)c_n(k) + \sum_{n=1}\sum_{m=0}^{n-1} V_n c_{n-m}^\dagger(k)c_{n+1+m}(k) + \text{H.C.}, \tag{4}$$

where subscript $n$ on $c(k)$ denotes the band index in the folded dispersion. The summation in the second term involves only odd band-index differences because we consider only forward–backward mode coupling, with no coupling between modes propagating in the same direction. The complicated coupling term in equation (4) can be illustrated directly in band structure (Fig. 1c) and schematically represented as a partially connected model (Fig. 2a) for forward/backward modes with different momenta. It is worth noting that the higher off-diagonal couplings (terms with $m \geq 1$ in (4)) cannot be neglected due to the relatively small frequency difference between neighboring bands in the moiré lattice. In contrast, the splitting is significant only near the degenerate point, since the bands span a large frequency range in general photonic/phononic crystals, which corresponds to a cutoff at $m = 0$.

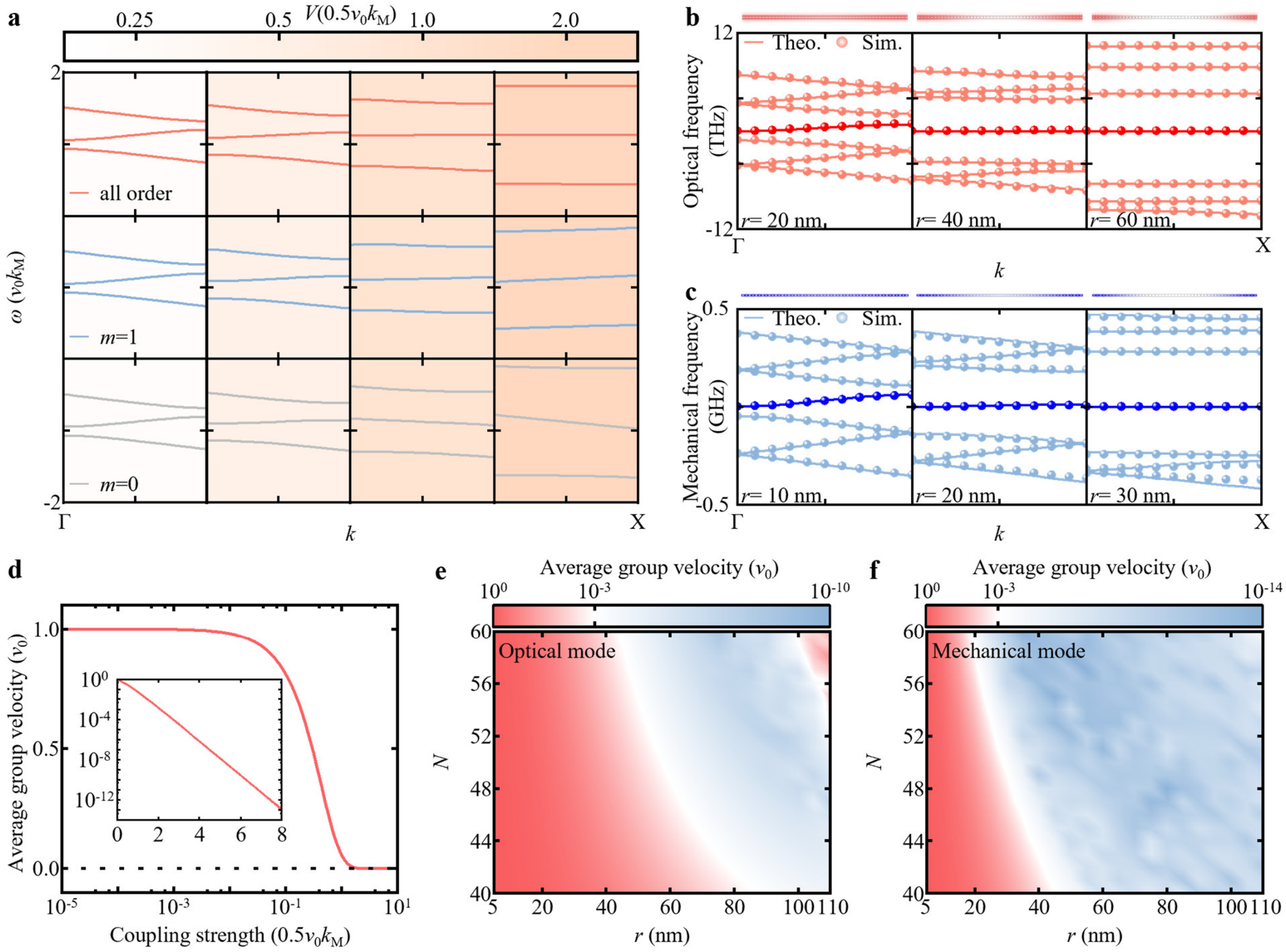

**Fig. 2. Effective Hamiltonian model for band calculation and comparison with simulation. a.** Comparison of band structures with different coupling strengths $V$ and neighboring order $m$. In the calculation, the initial group velocity $v_0$ is related to the intrinsic bandwidth $\Delta\omega_0=0.5v_0k_M$. At low coupling strengths ($V \le \Delta\omega_0$), band splitting is weak, and higher-order neighboring band couplings have less effect. At sufficiently strong coupling strengths ($V > \Delta\omega_0$), high-order neighboring band coupling leads to significant flattening of bands. **b** and **c.** Comparison of band structures obtained from finite element method (FEM) simulations and effective Hamiltonian calculations for optical and mechanical modes for different radii $r$ with $N = 50$. The optical and mechanical bands of interest (the 51st band) are highlighted in bright red and dark blue with the corresponding

electric field distribution $|\mathbf{E}|$ and displacement field distribution $|\mathbf{u}|$ shown on top of the band structure. The red and blue lines are the results calculated by effective Hamiltonian while the dots are the results obtained by FEM simulations. The frequencies of the optical and mechanical modes are shifted relative to the band of interest. **d.** Average group velocity for the 51st band calculated by effective Hamiltonian model, where $V$ is the same nonzero value of $V_n$ for $n$=50 and 51, which corresponds to the second band in a. Overview: The range of coupling strength varies largely from $10^{-5}$ to 10. Inset: Average group velocity decreases nearly exponentially for coupling strength from 0 to 8 (in units of: $0.5v_0k_M$). **e**. and **f.** Average group velocity for the $51^{st}$ optical band and mechanical band for one-dimensional moiré optomechanical crystal with different $N$ and radius of air holes $r$, which correspond to the highlighted bands in b and c. The simulation is carried out using 2D FEM simulation and group velocity is normalized by the initial group velocity $v_0$ of the nanobeam without modulation of air holes. All the FEM simulations are performed using 2D FEM simulation, with geometric parameters $a_1$ =508.47 nm, $a_2$=518.64 nm, $a_M$ =25.932 μm, and the width of nanobeam $w$=406.78 nm.

In order to study the flat-band formation process under different coupling parameters, we numerically extract the eigenvalues (Fig. 2a) for the effective Hamiltonian (4) under different coupling strengths $V$ with cutoffs on $m$ up to 0, 1 (up to the nearest and next-nearest neighboring band coupling), and without cutoffs. For simplicity, we assume equal coupling strengths for orders $n = 50, 51$ and analyze three bands ($50^{th}$ to $52^{nd}$) under different coupling strengths. From Fig. 2a, at low coupling strength $V < \Delta\omega_0 = \frac{1}{2}v_0 k_M$, band splitting exhibits little dependence on the cutoff order, since higher-order neighboring band splitting is relatively negligible. In contrast, at higher coupling strengths $V \geq \Delta\omega_0$ , bands with higher-order neighboring band coupling become markedly flatter, whereas those with a limited cutoff on $m$ still produce dispersive bands.

First, these results demonstrate that higher-order neighboring band couplings (the multi-inter-band coupling) play an essential role in flat-band formation. Due to the small frequency differences, higher-order neighboring band couplings make multiple bands couple together into a complex connected network. Strong inter-band couplings dominate the intrinsic bandwidth of the system, effectively suppressing initial dispersion and reducing the wave-vector dependence of the band frequencies, i.e., forming flat bands, where the dispersions are anchored by the neighboring bands through strong inter-band hybridization. Thus, the interplay of both higher-order neighboring band couplings and strong coupling strength is key to achieving flat bands (Fig. 2a).

Then, we compare the band structures from finite element method (FEM) simulation and effective Hamiltonian calculation to confirm the applicability of the effective Hamiltonian model. We take the 51st band for both optical and mechanical modes as an example highlighted in Fig. 2b and Fig. 2c by bright red and blue colors respectively. We find that the bands of interest calculated from the effective Hamiltonian agree with the results from FEM simulation under different radii of air holes. Moreover, the features of neighboring bands can also be described by the effective Hamiltonian models which are shown in red and blue in Fig. 2b and Fig. 2c. These results show the effective Hamiltonian can capture and describe the multi-band structure and the evolution from dispersive to flat bands in the proposed moiré optomechanical crystal. The above comparison between the effective Hamiltonian and the FEM simulation also supports the practical numerical condition $V \geq \Delta\omega_0$ for flat-band occurrence. This is a practical crossover rather than magic configurations: as the coupling strength increases, the bands become progressively flatter. More details about the comparison between the effective Hamiltonian and the FEM simulations are provided in the Supplementary Information.

Compared to multilayer systems[8, 9], in the proposed moiré optomechanical crystal, although modes selectively couple due to quasi-momentum conservation in the minimized Brillouin zone, which is characteristic of moiré systems, differences lie in the coupling mechanism: in our system, coupling arises from diffractive modulation on a single layer rather than interlayer tunneling. In those multilayer systems, interlayer tunneling strength and geometric parameters (such as twist angles or inter-layer distance) act as dimensionless perturbations to the renormalized group velocity, which undergoes repeated zero crossings as the perturbation grows[8, 9], producing alternating flat and non-flat transitions despite high coupling strength. This group velocity fluctuation is comparable to the initial group velocity[8], so high-quality confinement requires specific magic configurations. As these zero crossings are field-specific, simultaneously balancing them for distinct physical fields is difficult.

To further compare the parametric dependence of group velocity of the proposed moiré optomechanical crystal with that of multilayer systems, we numerically calculate the average group velocity $\bar{v}$ (which is defined as the arithmetic average of the absolute group velocity) as a function of diffractive coupling strength $V$ with the effective Hamiltonian model (Fig. 2d). In this calculation, we consider the simplest model with the same nonzero values of $V_n$ for $n$=50 and 51. From numerical calculation in Fig. 2d, increasing $V$ leads to a monotonic, nearly exponential decrease in group velocity $\bar{v}$ without crossing the zero-point. Specifically, we numerically find that when $V \geq \Delta\omega_0 = \frac{1}{2} v_0 k_{\mathrm{M}}$, the group velocity drops to approximately 1% or less of the initial group velocity, which can be regarded as a rough condition for flat-band formation. When $V$ is on the order of several $\Delta\omega_0$, the average group velocity decreases from $10^{-2}$ to $10^{-12}$ of the initial group velocity $v_0$, which strongly suppresses inter-cell energy transport in this propagating direction. Importantly, this monotonic, one-way suppression of the group velocity is the key advantage: stronger coupling makes the bands flatter, so the same structural parameters simultaneously improve confinement for different kinds of physical fields without balancing magic configurations.

To confirm the theoretical result with the FEM simulation, we further carried out a two-dimensional FEM simulation for the 51$^{\text{st}}$ band for optical (Fig. 2e) and mechanical modes (Fig. 2f) varying $N$ and $r$ with the other geometrical structure parameters the same as in Fig. 3a. From Fig. 2e and 2f, we observe a consistent trend: average group velocity decreases approximately exponentially with increasing coupling strength (roughly equivalent to larger air holes). For both optical and mechanical modes, increasing the radius of the air holes induces a decrease in the average group velocity in most of the parametric region, without repeatedly crossing the zero point of the group velocity. In Fig. 2e, with large $N$ and large $r$, it seems the average group velocity of the optical mode increases with $r$, breaking the above theoretical result. However, this increase in average group velocity arises from the coupling-induced splitting between the flat-band mode and the lossy mode propagating mostly in air due to the tiny tip structures in this parametric region, which is discussed further in the Supplementary Information. Apart from this parameter region, the same structural parameters flatten both optical and mechanical bands, confirming the field-independent character of the mechanism.

Therefore, both theory and simulation converge on the same central result: the average group velocity decreases nearly exponentially with coupling strength, without the magic-configuration-like non-monotonic behavior. This monotonic, magic-angle-free suppression is the key to simultaneous multi-field flat bands: it removes the need to balance separate magic configurations for each kind of fields. Instead, choosing a set of geometric parameters with strong coupling ensures extremely low group velocities for all kinds of fields. From this perspective, superimposing modulation patterns onto the same structure to form a moiré lattice not only simplifies the fabrication but also enables simultaneous manipulation of different kinds of physical fields within the same moiré lattice. Similar behavior has been reported in two-dimensional moiré photonic cavities, where confinement can improve monotonically as the twist angle decreases[17].

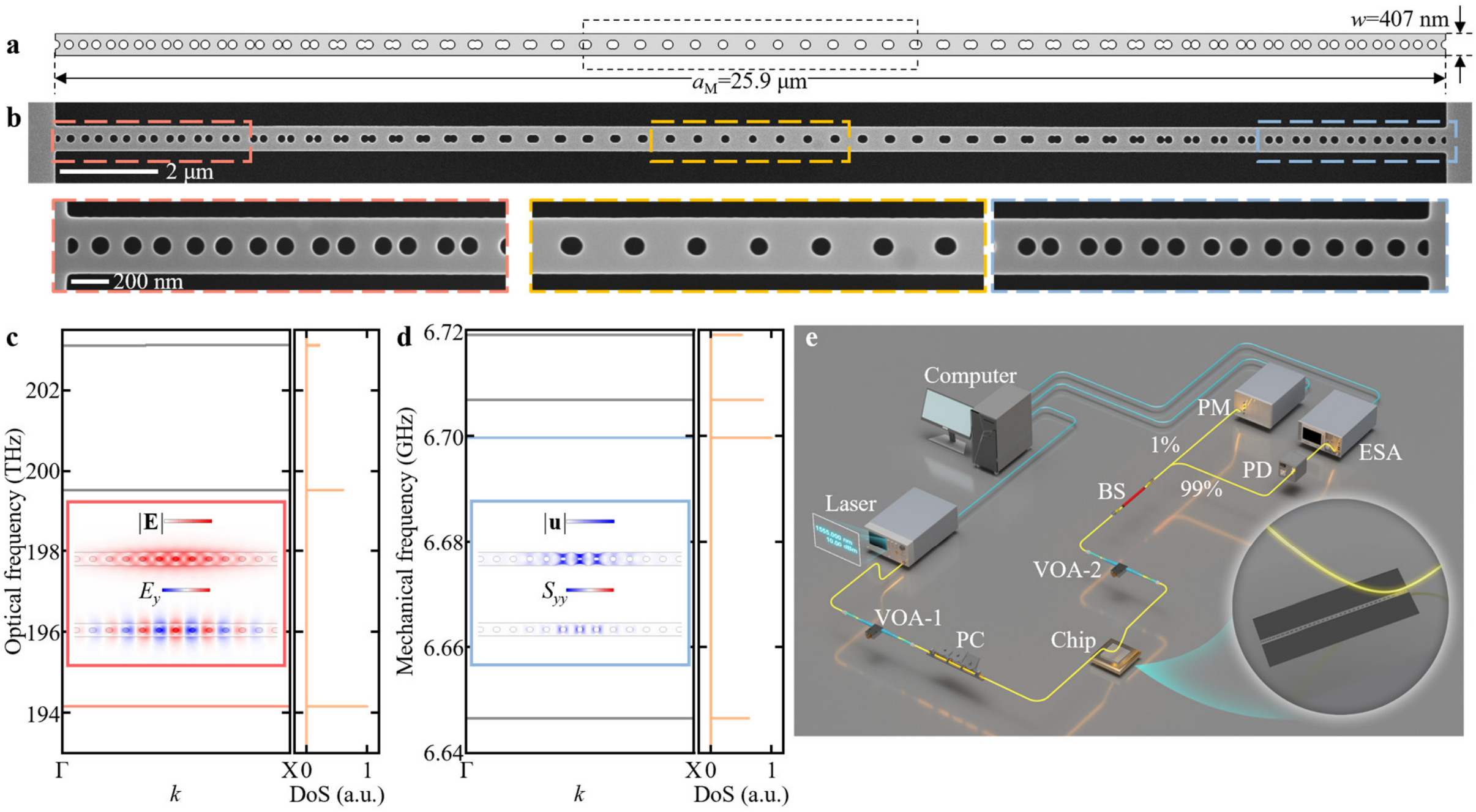

**Fig. 3. The designed moiré optomechanical crystal and experimental setup. a.** The moiré optomechanical crystal for simulation, with geometric parameters $r$ =81.356 nm, $a_1$ =508.47 nm, $a_2$ =518.64 nm, $w$ =406.78 nm and $a_M$ =25.932 μm. The height of the nanobeam is chosen as $h$ =220 nm. The refractive indices for silicon and air background are $n_{silicon}$ =3.48 and $n_{air}$ =1, respectively. The density and elasticity matrix for the silicon material are $\rho$=2330 kg/m$^3$ and ($D_{11}$, $D_{12}$, $D_{22}$, $D_{13}$, $D_{23}$, $D_{33}$, $D_{44}$, $D_{55}$, $D_{66}$) = (166, 64, 166, 64, 64, 166, 80, 80, 80) GPa

for nonzero terms. **b.** SEM images of the fabricated one-dimensional moiré optomechanical cavity. Top: Overview of the one-dimensional moiré optomechanical cavity. Bottom: Magnified views of the regions marked by the dashed boxes on the top, with colors matching those in the overview. **c.** Optical band structure of the moiré optomechanical crystal, with the targeted flat band highlighted in red. Inset: Spatial field distribution of the optical mode corresponding to the flat band of interest. **d.** Mechanical band structure of the moiré optomechanical crystal, with the targeted flat band highlighted in blue. Inset: Spatial field distribution of the mechanical mode corresponding to the flat band of interest. All the simulations in c and d are performed using 3D FEM simulation. **e.** Schematic of the experimental setup. Abbreviations: VOA-1/2: variable optical attenuator; PC: polarization controller; BS: 99:1 fiber beam splitter; PM: power meter; PD: high-speed photodetector; ESA: electrical spectrum analyzer.

The preceding analysis shows that flat-band formation is governed by the common mechanism applicable to the optical and mechanical modes. Based on the preceding analysis, we identify a set of structural parameters (Fig. 3a) for which multiple flat bands for both optical and mechanical fields simultaneously emerge as shown in Fig. 3c and Fig. 3d. These flat bands exhibit strong suppression of group velocity along the propagation direction, leading to mode localization. Other flat bands correspond to other localized states with distinct symmetry or spatial field distribution. As shown in Fig. 3c and Fig. 3d, the corresponding optical (194.15 THz) and mechanical (6.700 GHz) modes are strongly localized near the center of the unit cell illustrated in Fig. 3a. The high density of states (DoS) associated with flat bands provides the possibility for enhancing light–matter interactions, which is essential for cavity optomechanics[37].

The one-dimensional moiré optomechanical crystal allows for the construction of a long suspended nanobeam containing several moiré periods, such that periodic condition is approximately maintained near the center of the structure, where the flat-band modes can be localized. However, due to the near-zero group velocity, energy leakage to the adjacent unit cell is intrinsically suppressed, enabling strong confinement even with only a few unit cells—or, in the limiting case, a single unit cell. Motivated by this, we design a one-dimensional moiré optomechanical cavity using only a single unit cell of the one-dimensional moiré optomechanical crystal that supports both flat-band modes localized at its center. We simulate the configuration with three unit cells as a comparison and find that the performance for a single-unit-cell cavity is not significantly degraded (details in Supplementary Information).

Simulations of the cavity confirm that the resonant frequencies and mode distributions of both the optical and mechanical cavity modes closely match those of the flat band modes obtained under periodic boundary conditions. Specifically, the corresponding optical cavity mode resides at 194.15 THz with a simulated quality factor of $4.99\times10^5$, while the mechanical cavity mode appears at 6.700 GHz. In order to confirm the interaction between flat-band optical and mechanical modes, we calculate the optomechanical coupling rate $g_0$, which comprises contributions from both the photoelastic effect $g_{\mathrm{pe}}$ and the moving boundary effect $g_{\mathrm{mb}}$. In this cavity, $g_{\mathrm{pe}}/2\pi$ is 0.34 MHz, mainly due to the overlap between the optical electric field $E_y$ and the mechanical strain field $S_{yy}$, and $g_{\mathrm{mb}}/2\pi$ is 0.22 MHz, yielding a total optomechanical coupling rate $g_0/2\pi$ of 0.56 MHz. Other flat-band modes can also be localized within the cavity, but due to the different spatial distribution, the corresponding optomechanical coupling rate is smaller, making it difficult to probe the corresponding mechanical modes via cavity optomechanics. Thus, we mainly use the above-mentioned optical and mechanical modes to confirm the localization and interaction between optical and mechanical flat-band modes in the moiré lattice. This spatial overlap is a direct consequence of flat-band localization, rather than of a defect-defined cavity mode.

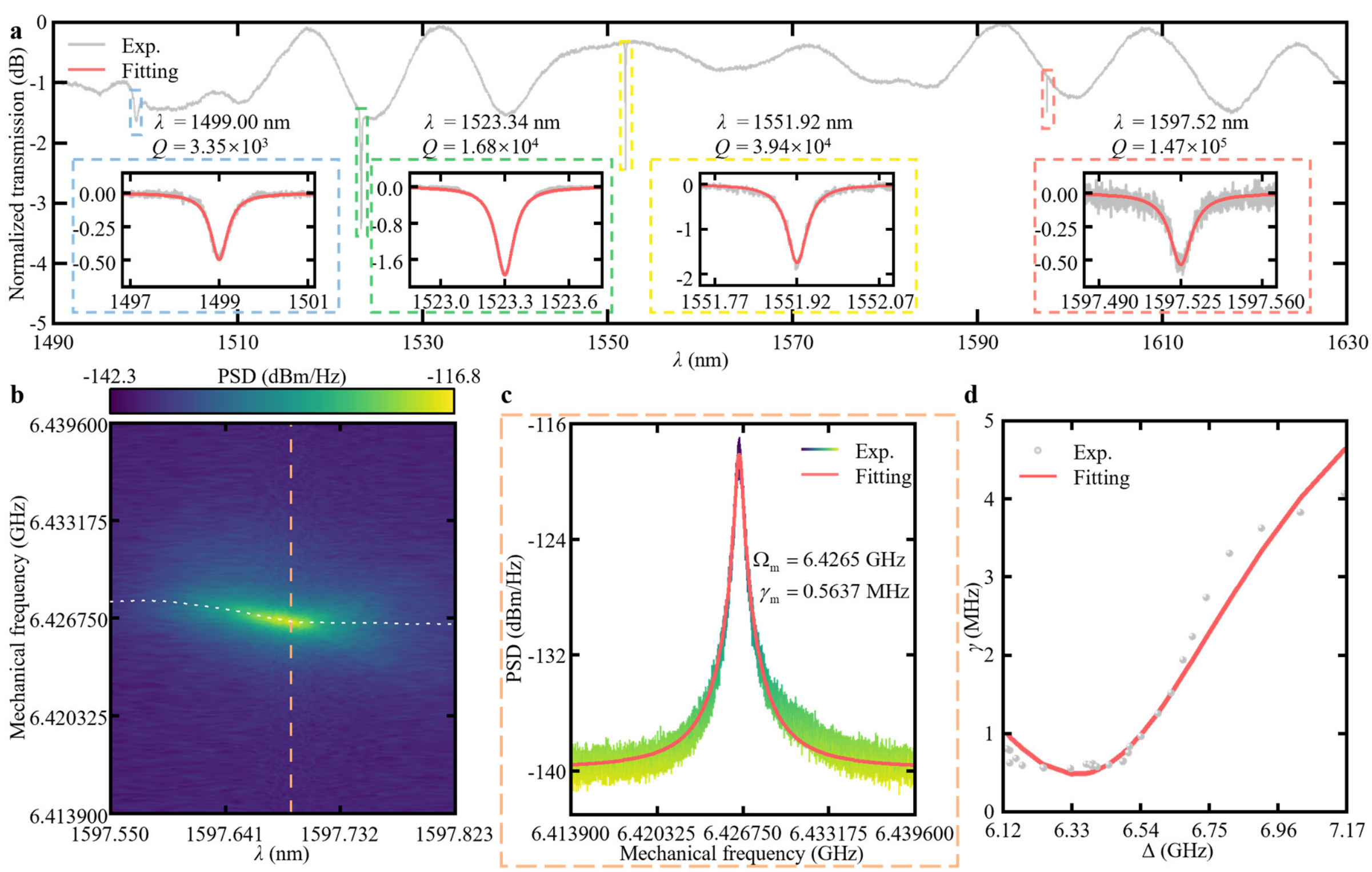

**Fig. 4. Characterization of the optical and mechanical modes. a.** Measured optical transmission from 1490 nm to 1630 nm with a step of 1 pm, where four dips can be observed in the overview. Insets: The characterization of the four optical modes. For the modes centered at 1499.00 nm and 1523.34 nm, scans with a step of 1 pm are performed; for the modes centered at 1551.92 nm and 1597.52 nm, fine scans with a step of 0.1 pm are performed. The colored dashed boxes correspond to those in the overview, where the red solid lines represent fits to the transmission spectra. **b.** Measured mechanical power spectral density (PSD) at different laser wavelengths with an input power of 10.69 dBm. The white dashed line corresponds to the resonance frequency of mechanical modes. **c.** PSD at the wavelength of 1597.6930 nm (purple dashed line in b) shows the mechanical resonance at 6.4265 GHz with an effective linewidth of 0.5637 MHz. **d.** Mechanical dissipation rate of the measured mode as a function of detuning.

## 3. Characterization of one-dimensional moiré optomechanical cavity

The designed one-dimensional moiré optomechanical cavity was fabricated on a silicon-on-insulator (SOI) chip. The device pattern was first defined on an electron beam resist layer using electron beam lithography (EBL), and then transferred to the top silicon layer via inductively coupled plasma (ICP) etching. The buried silicon dioxide layer was removed by buffered hydrofluoric acid (BHF) etching to release the suspended nanobeam, followed by critical point drying to minimize surface-tension effects. The scanning electron microscopy (SEM) image of the fabricated device is shown in Fig. 3b, in agreement with the design shown in Fig. 3a.

The experimental setup for characterization is illustrated in Fig. 3e. A pump laser is sent through a variable optical attenuator (VOA-1), with polarization adjusted by a polarization controller (PC). The light is then coupled into the cavity via a tapered optical fiber, exciting the optical cavity mode at frequency $\omega_{\mathrm{cav}}$. Due to optomechanical coupling, the transmitted optical signal carries the modulation induced by the mechanical mode. This output signal from the taper fiber passes through the second attenuator (VOA-2) and is then divided by a 1:99 fiber beam splitter. The 1% branch is monitored by a power meter (PM), while the remaining 99% is directed to a high-speed photodetector (PD). The detected electrical signal is analyzed by an electrical spectrum analyzer (ESA) to extract the mechanical power spectrum.

Based on the experimental platform described above, the optical transmission spectrum is presented in Fig. 4a. Four main dips are observed in the figure, whose frequency separations are consistent with the ones for four flat bands. We characterize the optical modes in boxed regions which are shown in the insets of Fig. 4a. By fitting, the four observed optical resonances within the measured wavelength range are located at 1597.52 nm, 1551.92 nm, 1523.34 nm and 1499.00 nm with intrinsic quality factors of $1.47\times10^5$, $3.94\times10^4$, $1.68\times10^4$ and $3.35\times10^3$, respectively, confirming multiple optical flat-band modes in this moiré optomechanical crystal. The resonance at 1597.52 nm corresponds to the optical flat-band mode identified in Fig. 3c, in which the frequency deviation from the designed value arises primarily from fabrication-induced variations. The observation of four well-resolved resonances is consistent with the presence of multiple flat-band modes surviving fabrication disorder.

The power spectral density (PSD) of the electrical signal collected from the photodetector reflects the mechanical mode dynamics when the optical mode is probed at the detuned wavelength of 1597.6930 nm (with the optical resonance at 1597.52 nm at low output laser power), as shown in Fig. 4b and Fig. 4c. The PSD contains contributions from the mechanical resonance spectrum $S_{PP}(\Omega)$, modulated by the transduction function $H(\Delta,\Omega)$ of the system[44, 45, 54]. By fitting the measured PSD with the transduction model, the optical detuning can be extracted. Due to the optomechanical interaction, the mechanical frequency and dissipation rate are modified[31, 55]. Fitting the mechanical linewidth using these relations yields an intrinsic mechanical frequency of 6.4169 GHz and an intrinsic mechanical damping rate of 6.9679 MHz, corresponding to an intrinsic mechanical quality factor $Q_{\mathrm{m,intrinsic}} \approx 9.21\times10^{2}$. Also, the optomechanical coupling rate is extracted to be 0.3874 MHz from Fig. 4d. We also detect another mechanical flat-band mode in our cavity with the same optical modes, which is shown in the Supplementary Information and confirms multiple mechanical flat-band modes in the one-dimensional moiré optomechanical cavity. The extracted coupling rate is consistent with the simulated value, indicating that the flat-band mode and its optomechanical overlap are largely preserved.

From the above results, we confirm that the one-dimensional moiré optomechanical crystal supports multiple co-localized optical and mechanical modes arising from flat bands, in agreement with numerical predictions. Despite fabrication-induced imperfections, we observe a high-$Q$ optical mode with quality factor up to $1.47\times10^{5}$, which is of the same order of magnitude as the simulation value. In parallel, we detect a mechanical resonance with an intrinsic quality factor of approximately $9.21 \times 10^{2}$, together with its modulation through optical detuning, corresponding to an optomechanical coupling rate of about 0.3874 MHz. These results demonstrate the simultaneous confinement of both optical and mechanical modes as well as their optomechanical interaction within a moiré optomechanical cavity—achieved without resorting to conventional bandgap or defect engineering, highlighting the potential of moiré engineering as an alternative platform for optomechanics.

## 4. Discussion and Conclusion

In this work, we have introduced, to the best of our knowledge, the first optomechanical crystal cavity based on a one-dimensional moiré lattice. We developed an effective Hamiltonian model for the moiré lattice formed by superimposing patterns within the same structure, which captures the mechanism of flat-band formation and agrees with the FEM simulations. Unlike multilayer moiré systems, the effective Hamiltonian model shows that flat-band formation in a single-layer moiré system arises from multi-band coupling and leads to a monotonic, exponential decrease in group velocity, independent of the field types. The group velocity in our moiré systems decreases exponentially as coupling strength increases without inducing undesired group velocity fluctuations from repeated crossings of the zero-velocity points[8, 9, 21]. This is crucial for the simultaneous confinement of different types of fields, since balancing distinct zero-velocity points for each field is challenging.

In the proposed moiré optomechanical crystal, multiple optical and mechanical flat-band modes can be simultaneously co-localized and interact. Experimentally, we observe quality factors of $1.47\times10^5$ and $9.21 \times 10^2$ for the optical and mechanical modes, respectively, together with an optomechanical coupling rate $g_0 / 2\pi$ of 0.3874 MHz. These results demonstrate that flat-band modes can simultaneously confine photons and phonons and mediate their coupling, which is the essential functionality for cavity optomechanics. Importantly, our approach establishes a new design paradigm for optomechanical cavities—one that avoids the traditional reliance on bandgap or defect engineering.

Looking ahead, the performance of the moiré optomechanical cavity can be further optimized through two complementary strategies: (1) independently designing the modulation patterns of the original lattices to precisely control the band structures and mode profiles, and (2) fine-tuning the local pattern within the central localization region[18] to enhance spatial overlap between optical and mechanical modes, thereby increasing the optomechanical coupling strength. Since the manipulation of different types of fields may be achieved on the same moiré lattice, this framework can be naturally extended to investigate further multi-field interactions, such as acoustics–fluid mechanics[49], phononics–heat transport[50], and light–matter coupling[51]. Moreover, the framework is not limited to one-dimensional systems; extending it to higher-dimensional systems could be explored, for example, in two-dimensional cavity optomechanics, where increased thermal conductance[56] may further improve the device performance. Importantly, by decoupling the design of mode localization (via the moiré lattice) from that of mode coupling (via local pattern engineering), this approach may offer a route to devices that simultaneously achieve high quality factors and strong coupling rates, which is essential for exploring strong-coupling phenomena, quantum applications with high quantum cooperativity[57] and so on.

In conclusion, we have demonstrated multiple flat bands for multiple types of physical fields in a single-layer moiré system. The underlying mechanism, strong inter-band coupling in single-layer moiré lattices that monotonically suppresses the group velocity without repeated zero crossings, is universal across different physical fields and removes the need for magic-configuration tuning. Using this mechanism, we realized a new class of optomechanical cavity based on flat-band physics in a one-dimensional moiré optomechanical crystal. This platform not only broadens the design space of cavity optomechanics but also establishes a field-independent, magic-configuration-free design principle for multi-field localization and interaction in compact hybrid moiré systems.

**Acknowledgements:** The authors express their gratitude to Tianjin H-Chip Technology Group Corporation, Innovation Center of Advanced Optoelectronic Chip and Institute for Electronics and Information Technology in Tianjin, Tsinghua University for their fabrication support of the device.

**Funding:** This research was funded by the National Natural Science Foundation of China (Grant No. 32541100, No. U22A6004); Beijing Frontier Science Center for Quantum Information; and Beijing Academy of Quantum Information Sciences.

**Author contributions:** Z.C., N.W. and K.C. conceived the study. K.C. supervised the project. K.C., Y.H. and N.W. advised on the device optimization. Z.C. performed the theoretical analysis and designed the device. Z.C. and S.L. performed the band and mode field simulations. Z.C. and C.W. conducted the experiments. Z.C. analyzed the experiment data. Z.C. and K.C. wrote the paper. All authors discussed the results and implications throughout the investigation.

**Competing interests:** The authors declare no conflicts of interest.

**Data availability:** All data needed to evaluate the conclusions in the paper are present in the paper and/or the Supplementary Information.